\documentclass[lettersize,journal]{IEEEtran}
\usepackage{amsmath,amsfonts}
\usepackage{array}
\usepackage[caption=false,font=normalsize,labelfont=sf,textfont=sf]{subfig}
\usepackage{textcomp}
\usepackage{stfloats}
\usepackage{url}
\usepackage{verbatim}
\usepackage{graphicx}
\usepackage{algorithm}
\usepackage{algpseudocode}
\usepackage{xcolor}
\usepackage{makecell}
\usepackage{fancyvrb}
\usepackage{lineno}
\usepackage{etoolbox}
\usepackage{arydshln}
\usepackage{booktabs}  
\usepackage{threeparttable} 
\usepackage{titlesec}
\usepackage{amsmath}
\usepackage{multirow}
\usepackage[colorlinks=true, linkcolor=blue, citecolor=blue]{hyperref}

\usepackage{tikz,xcolor,hyperref} 
\definecolor{lime}{HTML}{A6CE39}
\DeclareRobustCommand{\orcidicon}{
\begin{tikzpicture}
\draw[lime, fill=lime] (0,0) 
circle [radius=0.16] 
node[white] {{\fontfamily{qag}\selectfont \tiny ID}};    \draw[white, fill=white] (-0.0625,0.095) 
circle [radius=0.007];    \end{tikzpicture}
\hspace{-2mm}}
\foreach \x in {A, ..., Z}{
\expandafter\xdef\csname orcid\x\endcsname{\noexpand\href{https://orcid.org/\csname orcidauthor\x\endcsname}{\noexpand\orcidicon}}}

\begin{document}

\title{An Improved Two-Step Attack on Lattice-Based Cryptography: A Case Study of Kyber}

\author{Kai Wang\orcidB{}, Dejun Xu\orcidA{}, and Jing Tian\orcidC{}, \textit{Member, IEEE}

\thanks{
This work was supported in part by the Natural Science Foundation of Jiangsu Province of China under Grant BK20243038, in part by the Key Research Plan of Jiangsu Province of China under Grant BE2022098, and in part by the Young Elite Scientists Sponsorship Program by the China Association for Science and Technology under Grant 2023QNRC001. \textit{(Corresponding author: Jing Tian.)}}

\thanks{The authors are with the School of Integrated Circuits, Nanjing University, Suzhou, 215163, China (e-mail: tianjing@nju.edu.cn).}
}

\markboth{IEEE Transactions on Computer-Aided Design of Integrated Circuits and Systems}{Wang \MakeLowercase{\textit{et al.}}: An Improved Two-Step Attack on Lattice-Based Cryptography: A Case Study of Kyber} 

\maketitle

\begin{abstract}
After three rounds of post-quantum cryptography (PQC) strict evaluations conducted by NIST, CRYSTALS-Kyber was successfully selected in July 2022 and standardized in August 2024. It becomes urgent to further evaluate Kyber's physical security for the upcoming deployment phase. In this brief, we present an improved two-step attack on Kyber to quickly recover the full secret key, $\mathbf{s}$, by using much fewer power traces and less time. In the first step, we use the correlation power analysis (CPA) to obtain a portion of guess values of $\mathbf{s}$ with a small number of power traces. The CPA is enhanced by utilizing both Pearson and Kendall's rank correlation coefficients and modifying the leakage model to improve the accuracy. In the second step, we adopt the lattice attack to recover $\mathbf{s}$ based on the results of CPA. The success rate is largely built up by constructing a trial-and-error method. We deploy the reference implementations of Kyber-512, -768, and -1024 on an ARM Cortex-M4 target board and successfully recover $\mathbf{s}$ in approximately $9\sim 10$ minutes with at most 15 power traces, using a Xeon Gold 6342-equipped machine for the attack.
\end{abstract}


\section{Introduction}
\IEEEPARstart{W}{ITH} the fast development of quantum computing, modern cryptographic algorithms are encountering significant threats. Recognizing this problem, NIST started the  post-quantum cryptography (PQC) standardization process in 2016 with the aim of standardizing quantum-resistant cryptographic algorithms. By July 2022, NIST released the post-quantum cryptographic standard candidates including three signatures and one key encapsulation mechanism (KEM) algorithm after the third round evaluation~\cite{alagic2022status}. CRYSTALS-Kyber is the only KEM, which has been successfully standardized as module-lattice-based KEM (ML-KEM) noted as FIPS 203 from August 2024. The mathematical security of Kyber is widely recognized, but its physical security demands urgent attention in the deployment phase~\cite{ravi2024side,zhao2024tpe}. 

In recent years, several side-channel attacks (SCAs) on Kyber~\cite{ravi2020generic,ueno2022curse,shen2023find,yang2023chosen,kuo2023lattice} have been proposed by researchers. In 2020, Ravi \textit{et al.}~\cite{ravi2020generic} used plaintext-checking (PC) oracle-based SCAs to recover the secret key of Kyber-512 with 2560 power traces according to the vulnerabilities in Fujisaki-Okamoto (FO) transform for the first time. Then for correlation power analysis (CPA), Ueno \textit{et al.}~\cite{ueno2022curse} proposed the side-channel leakage in re-encryption as a PC oracle, utilizing 3072 power traces to recover the key. The work in~\cite{shen2023find} deals with imperfect PC oracle SCAs by using the CPA attack twice to recover 60\% secret key of Kyber-512 at the cost of about 1619 power traces, but there are still 40\% error coefficients. Contrast to previous works, in~\cite{yang2023chosen}, Yang \textit{et al.} tried to carry out a random ciphertext CPA attack on the reference implementation of Kyber-512, successfully recovering two coefficients of secret key using $20$ or more power traces. Nevertheless, they had to repeat the attack 256 times to recover the entire secret key of Kyber-512. So, they at least need 5120 power traces for a key. In~\cite{kuo2023lattice}, Yen-Ting Kuo and Atsushi Takayasu presented a two-step attack on Kyber by integrating the random ciphertext CPA attack with the lattice attack. They adopted the CPA attack to recover part of the coefficients of a key and then directly solved the rest coefficients by using a low-dimension equation of the M-LWE problem. The correctness was verified by using computer simulations. In this brief, we take the idea of~\cite{kuo2023lattice} as the starting point and develop an improved two-step attack method on Kyber to further validate it in practice and improve its efficiency.

The proposed two-step attack is summarized in Fig.~\ref{flow}. In the first step, we modify the leakage model to better align our CPA with the arithmetic logic unit. Additionally, we introduce an optional correlation coefficient, the Kendall rank correlation, to address the nonlinearity and robustness issues of the original CPA in real-world attacks. In the second step, we construct a trial-and-error lattice attack algorithm to recover the entire secret key and optimize the algorithm's parameters. By employing this two-step attack, we achieve significant improvements in both efficiency and accuracy in practice. In addition, the attack methods we proposed for the Point-Wise Multiplication (PWM) stage apply to several other lattice-based PQC. Dilithium employs the single-point PWM and is more vulnerable to similar attacks. The PWM of Falcon works with floating-point numbers, but significant side-channel leakage remains due to secret key and message multiplication~\cite{karabulut2021falcon}.

\section{Preliminaries}
\subsection{Kyber and Number Theoretic Transform in Kyber}\label{tab:paras_c}
Kyber consists of three stages: key generation, encryption, and decryption. In the key generation stage, the public key \(pk\) is constructed as $\mathbf{A}\mathbf{s} + \mathbf{e}$. During the encryption stage, the message $m$ is encrypted into ciphertext $c$. In the decryption stage, the receiver utilizes $\mathbf{s}$ to perform corresponding operations to recover the message $m$ from $c$. The main operations of decryption are as follows:
\begin{equation}
\label{1}
m\leftarrow Compress_q(v-INTT(\boldsymbol{\hat{s}}^T\circ NTT(\boldsymbol{u})),1).
\end{equation}
Computing polynomial multiplication in NTT domain can reduce the computational complexity from \(\mathcal{O}(N^2)\) to \(\mathcal{O}(N \log N)\), thus significantly improving the speed of the entire encryption and decryption process in Kyber. Note that for Kyber, here is only $256^{\text{th}}$ primitive root of unity $\xi$ instead of $512^{\text{th}}$. Therefore, 256 coefficients are divided into two groups, namely the odd and even groups, for NTT calculations. The even coefficients in NTT domain are described as follows:
\begin{equation}
\hat{f}_{2i} = \sum_{j=0}^{127}f_{2j}\xi^{(2br(i)+1)j} \Rightarrow \ [\hat{f}_{2i}]^{\top} = \mathbf{M} [f_{2i}]^{\top}, 0\leq i <128,
\end{equation}
where \text{$br$}($i$) represents bit-reversal of $i$ and $\mathbf{M}$ is an $128 \times 128$ matrix with reduced integers. Similarly, the inverse matrix of INTT can be represented as $\mathbf{N}$ in the same way.

\begin{figure}[!t]
\centering
\includegraphics[width=3.3in]{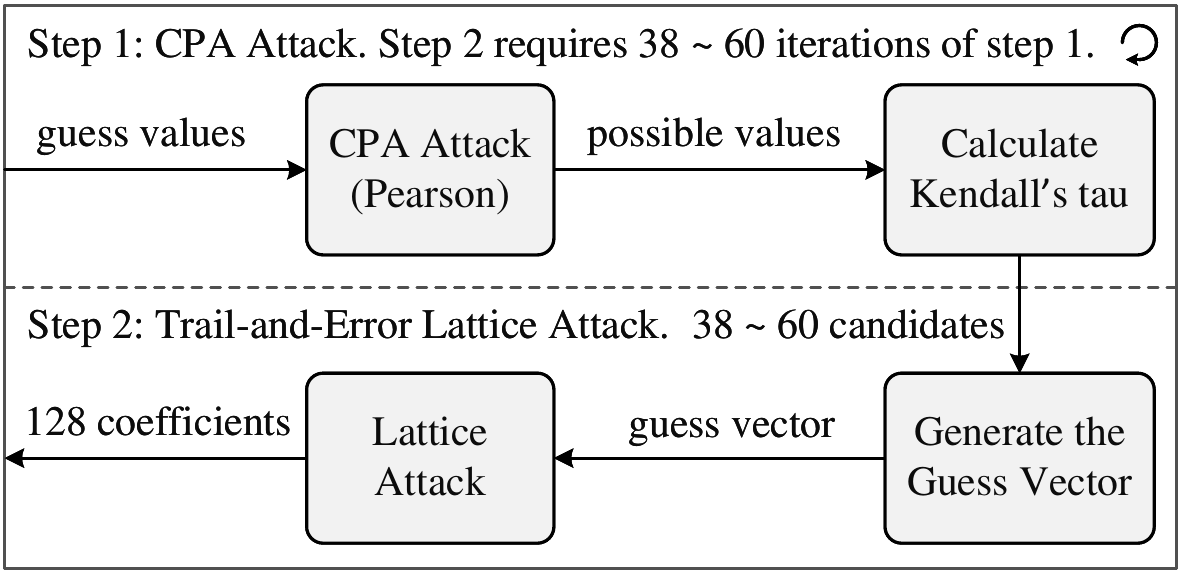}
\caption{The architecture of the proposed two-step attack.} 
\label{flow}
\end{figure}

\subsection{Correlation-Based Attack Techniques}
The conventional CPA attack is a non-invasive SCA that uses the Person correlation between power consumption models and actual power traces to recover cryptographic keys. It identifies the correct key by finding the highest correlation between measured and hypothetical values. Pearson correlation coefficient (PCC) is efficient and well-suited for detecting linear relationships between the hypothetical and actual values.

In contrast, Kendall's tau is a non-parametric measure of association that quantifies the degree of concordance between the rankings of two variables, indicating whether their values are consistently ranked together~\cite{abdi2007kendall}. Kendall's tau $\tau$ ranges from $-1$ to $1$, with $\tau=1$ indicating perfect agreement, $\tau=-1$ indicating perfect disagreement, and $\tau=0$ indicating no agreement. The Kendall's tau calculation is computed as follows:
\begin{equation}
\tau_b = \frac{c - d}{\sqrt{({c+d+t_x})({c+d+t_y})}},
\end{equation}
where $c$ and $d$ denote the numbers of concordant and discordant pairs, $t_x$ and $t_y$ represent the counts of tied ranks in the $x$ and $y$ data sets, respectively.

We have separately implemented the two methods in software and found that the PCC computation is much faster than the Kendall's tau calculation. The main reason is that the latter must compute a large full permutation. Luckily, when we combined the two methods together, the efficiency and reliability both were enhanced. More details will be shown in the following section.

\begin{figure}
\centering
\includegraphics[width=3.5in]{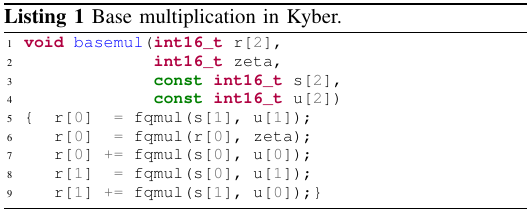}
\label{basemul}
\end{figure}

\section{Proposed Improved Two-Step Attack}\label{Sec:AM}
It should be remarked that the proposed improved two-step attack is illustrated in Fig. \ref{flow}. We assume that an attacker has access to a device which is running Kyber decryption and can only observe ciphertexts entered into the device. In addition, they can also capture the power traces of the device.
 
Several test vector leakage assessment (TVLA) results from previous works have indicated the potential leakage points during the decryption process in Kyber~\cite{shen2023find,yang2023chosen}. One critical side-channel leakage occurs during the PWM of $\hat{\mathbf{s}}\circ\hat{\mathbf{u}}$. In the clean implementation of pqm4~\cite{kannwischer2019pqm4}, polynomial multiplication employs 256 $basemuls$ in Kyber-512. Moreover, each calculation of $basemul$ is independent from the others. From Listing 1, it is clear that the result depends on two secret key coefficients, \( s_0 \) and \( s_1 \). All partial secret key coefficients range from $\left( 0, 3328 \right)$ for Kyber. In a practical attack, suppose that the attacker aims to recover the coefficient \( s_0 \). \( r_1 \) can be selected as the intermediate value.

\subsection{Modified Leakage Model for CPA}\label{UIHW}
\begin{figure}[!t]
\centering
\includegraphics[width=3.33in]{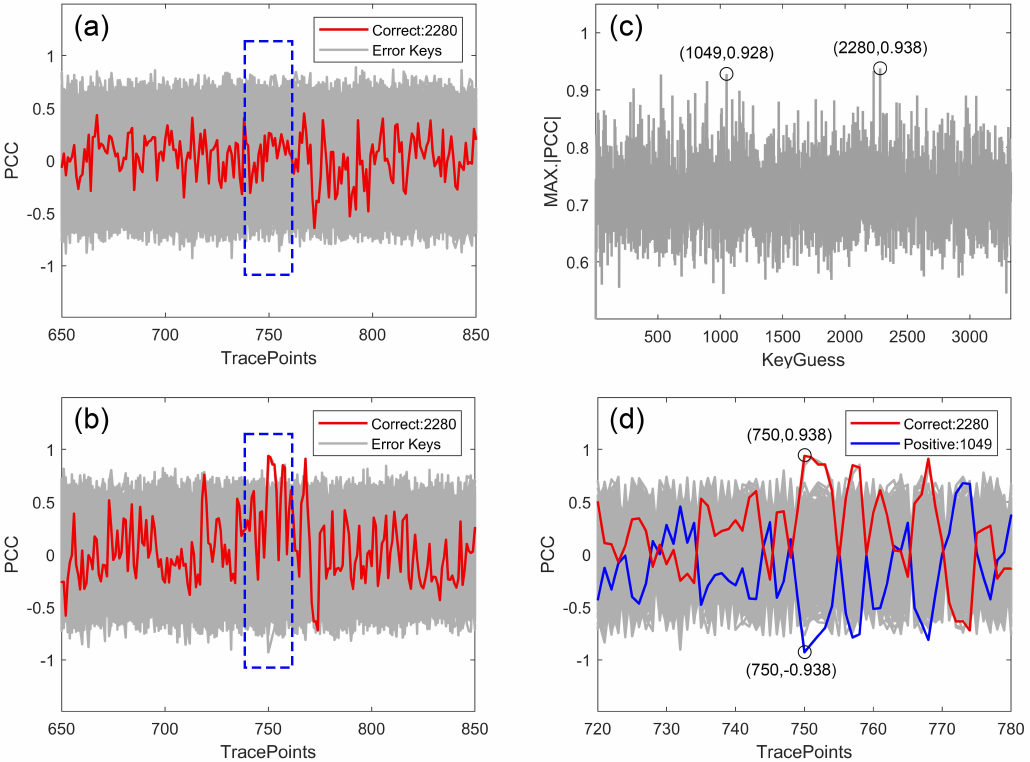}
\caption{(a) and (b) present the PCC values for all guessed keys in a $basemul$ call when \( D = 15 \), for the original and modified models, respectively. (c) shows the $|$PCC$|$ values for all guessed keys, while the blue curve in (d) illustrates the occurrence of false positives.} 
\label{fig_HW}
\end{figure}

The Hamming weight (HW) model counts the number of ``1'' in intermediate values and is a common linear power model. ARM processors' ALUs are optimized for two's complement operations, so we adjust the HW (called modified Hamming weight, MHW) to account for this by using the HW of the 16-bit complement of the intermediate values. This modification improves the accuracy of the HW model and reduces the number of power traces (denoted as $D$) needed for CPA. As shown inside the blue dotted box in Fig.~\ref{fig_HW} (a) and (b), the correct key coefficient is obscured using the original model (a) but can be effectively identified with just 15 power traces using the modified model (b).

\subsection{Optional Kendall’s tau for CPA}\label{AK}
When the number of power traces is limited, traditional CPA may face challenges due to high PCC values across guessed keys. To address this, we set the $|$PCC$|$ threshold (e.g., 0.9) based on repeated experimental analysis, with 0.9 performing well in most $basemul$ calls. We then map these candidates to a hypothetical power consumption matrix and compute Kendall's tau between this matrix and the actual power traces. The candidate with the highest Kendall's tau value is selected as the correct key coefficient, which helps to reduce false positives. This approach combines PCC for initial filtering with Kendall's tau for more precise selection, despite Kendall's tau being more computationally demanding.

\subsection{Trial-and-Error Lattice Attack after CPA}
As introduced in~\cite{kuo2023lattice}, suppose that an attacker has successfully recovered $sr$ coefficients out of a group of 128 secret key coefficients $\hat{\mathbf{s}}_i$, while the remaining $128-sr$ coefficients are unsuccessfully recovered. Let $I_a = (a_0, \ldots, a_{sr-1})$ denote the recovered coefficients indices, and $I_b = (b_0, \ldots, b_{127-sr})$  represent the indices of coefficients that have failed to be recovered, \textit{i.e.}, the unknown coefficients. INTT($\hat{\mathbf{s}}_i$) = $\mathbf{N}\hat{\mathbf{s}}_i$ = $\mathbf{s}_i$ mod $q$ can be rewritten as $\mathbf{N}_A\hat{\mathbf{s}}_{iA}$ + $\mathbf{N}_B\hat{\mathbf{s}}_{iB}$ = $\mathbf{s}_i$ mod $q$~\cite{kuo2023lattice}, where matrix $\mathbf{N}_A = [\mathbf{n}_{a_0}, \ldots, \mathbf{n}_{{a}_{sr-1}}]$ consists of columns in matrix $\mathbf{N}$ corresponding to the indices $I_c$, while $\hat{\mathbf{s}}_{iA} = [\hat{\mathbf{s}}_{a_0}, \ldots, \hat{\mathbf{s}}_{{a}_{sr-1}}]^{\top}$ represents the vector composed of successfully recovered coefficients. Similarly, $\mathbf{N}_B$ and $\hat{\mathbf{s}}_{iB}$ are obtained in the same manner. 

In the above formulas, both $\mathbf{N}_A$ and $\hat{\mathbf{s}}_{iA}$ are known. Let $\mathbf{t} = \mathbf{N}_A \hat{\mathbf{s}}_{iA}$, $\mathbf{A} = -\mathbf{N}_B$, and $\mathbf{s}' = \hat{\mathbf{s}}_{iB}$. Then, we obtain $\mathbf{t} = \mathbf{A} \cdot \mathbf{s}' + \mathbf{s}_i \mod q$. This conveniently forms a low-dimension LWE problem, which is simpler compared to the original problem in Kyber because the rank of $\mathbf{A}$ is smaller. We use Kannan's embedding technique to solve this problem~\cite{kannan1987minkowski}, where the matrix $\mathbf{B}_{kan}$ is summarized as follows: 
\begin{equation}
\mathbf{B}_{kan} = \left[\begin{array}{c;{1pt/2pt}c}
\mathbf{I}_{sr} \quad  \mathbf{A}' & \mathbf{0} \\ 
\quad \mathbf{0} \quad q\mathbf{I}_{n-sr} & \mathbf{0} \\ \hdashline[1pt/2pt]
\mathbf{t}^{\top} & 1
\end{array}\right].
\end{equation}
$\left[\mathbf{I}_{sr} \mid \mathbf{A}^{\prime}\right]$ is a reduced row echelon matrix of $\mathbf{A}$ transpose. A vector $\mathbf{w} = [\mathbf{s}_i^{\top} | 1]$ of length $129$ is obtained by reduction algorithm, which contains the secret key $\mathbf{s}_i^{\top}$ of length 128. For Kyber-512, when $sr \ge 39$, a lattice attack can recover 128 odd/even index coefficients. For Kyber-768/1024, when $sr \ge 38$, the corresponding coefficients can be correctly recovered. 

We elaborate on the theoretical foundation of lattice attacks, which can tolerate error coefficients of recovered key and still successfully retrieve all 128 key. Leveraging this method, we propose a more flexible and versatile approach for its application. The proposed trial-and-error lattice attack is shown in Listing 2, where $n$ represents the number of CPA, and $\mathbf{M}$ and $count$ denote the NTT linear matrix form of Kyber and the number of trials, respectively. In our method, we select 60 basemul calls for the CPA attack and use 38/39 of them as inputs for the lattice attack. If the attack fails, the smallest input is replaced with another CPA result. This process continues until the attack succeeds or all 60 results are used, representing a trade-off between computational complexity and success rate. After extensive experimentation, we found that using 60 results provides similar success to using 128. It is also important to note that as the dimension of the lattice problem increases, the computational complexity grows. For Kyber, 38/39 results are sufficient to handle the 128 coefficients of a key group. However, for Dilithium with a group size of 256, both the lattice dimension and the required number of CPA results will need to increase.

\begin{figure}
\centering
\includegraphics[width=3.5in]{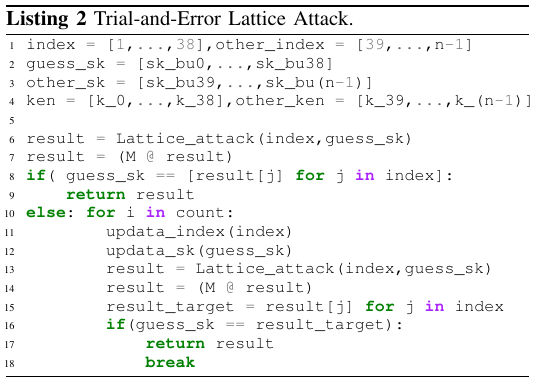}
\label{code:CPA+LA}
\end{figure}

\section{Experiments}\label{Sec:Exp}
We run the reference implementations of Kyber-512, -768, and -1024 from the pqm4 library \cite{kannwischer2019pqm4} with -o compilation optimization on a STM32F407-DISCOVERY target board at $48$ MHz. The power traces are captured using a PicoScope 3043D oscilloscope and a CYBERTEK EM5030-2 electromagnetic probe, with a $500$ MSa/s sampling rate. Based on the analysis in Section~\ref{Sec:AM}, we have chosen $fqmul(k, u_1)$ as the intermediate value of our attack.

\begin{figure}[!t]
\centering
\includegraphics[width=3.33in]{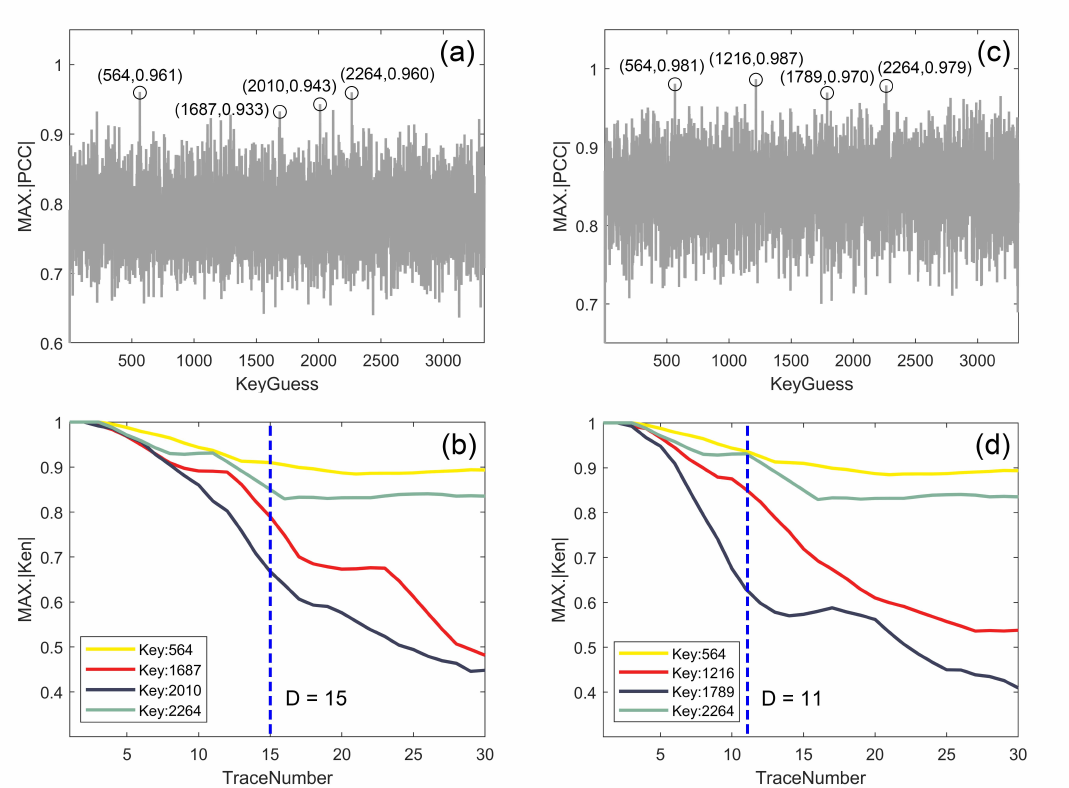}
\caption{(a) and (b) present the PCC and Ken results, respectively, for a another $basemul$ call when \( D = 15 \). Similarly, (c) and (d) show the PCC and Ken results for the same $basemul$ call when \( D = 11 \).} 
\label{fig_KEN}
\end{figure}

\subsubsection*{\textbf{Step1: CPA}}
As aforementioned, the CPA is divided into two stages: the capturing stage and the modeling computation stage. Our optimizations are mainly focused on the latter stage, including modifying the leakage model and adding an optional Kendall’s tau computing following the PCC computing.
 
For the capturing stage, we carry out the CPA at the NTT operation which calls $basemul$ functions 128 times. In our CPA experiments, we capture random ciphertexts and tremendous data of electromagnetic radiations during those calls. To show the effectiveness of the modified leakage model in the modeling computation stage, we have implemented two experiments. The former is shown in Fig.~\ref{fig_HW} (a) and (b) of Section~\ref{UIHW} and the modified model can reinforce the PCC value of the correct key. The latter shown in Fig.~\ref{fig_HW} (c) and (d) is to demonstrate its superiority in identifying the complementary false positive.

When the number of random ciphertexts $D$ is set to $15$, the relationship of the maximum $|$PCC$|$ to the guess keys is shown in Fig.~\ref{fig_HW} (c). We can directly pick out the correct key $2280$ and its complementary false positive $1049$ (equal to $3329-2280$), which is a general phenomenon. In addition, as the number of power traces decreases, the attacker would run into some problems because both the correct coefficient $s_{0}$ and its complementary value $q-s_{0}$  would have high and close PCC values, as the HW$(s_{0})$ and HW$(q-s_{0})$ are highly correlated. To escape such false positive, we have modified the leakage model as described in Section~\ref{UIHW}. The correspondence between the PCC values of the correct key coefficient and the false positive one is shown in Fig.~\ref{fig_HW} (d). It is observed that the positive peak value corresponds to the correct key coefficient and the negative peak value of PCC corresponds to the false positive one. Consequently, to avoid the false positive being selected, we should focus solely on the positive peak value of PCC. We briefly explain this phenomenon as follows: Let $s_0$ be the correct one, and $s_0' = 3329-s_0$ be the false positive one. The intermediate values for the two guessed key coefficients are summarized below:
\begin{equation}
r_1 = \text{fqmul}(s_0,u_1)=(s_0\times u_1) \, mod \, q, 
\end{equation}
\begin{equation}
\begin{aligned}
r_1' = \text{fqmul}(s_0',u_1)&=((3329-s_0)\times u_1) \, mod \, q \\
&=-(s_0\times u_1) \, mod \, q.
\end{aligned}
\end{equation}
MHW$(r_1)$ and MHW$(r_1')$ show a negative correlation, where the correct MHW$(r_1)$ is expected to be a positive PCC. Thus, considering their signs allows us to distinguish them.

\begin{figure}[!t]
\centering
\includegraphics[width=3in]{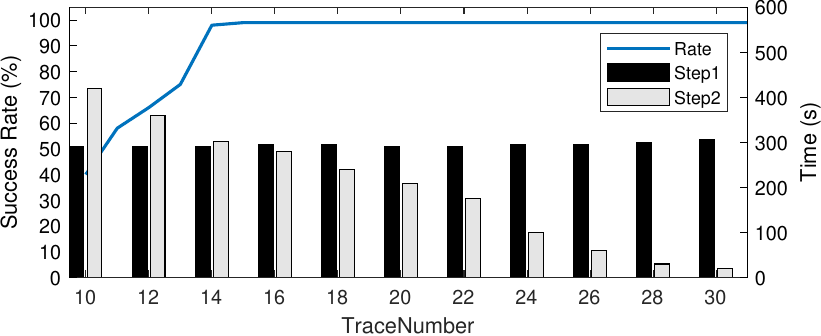}
\caption{The success rate and efficiency of our attack in twenty experiments.}
\label{fig_reslut}
\end{figure}

To further improve the accuracy in the CPA step, we adopt the Kendall’s tau to the picked candidates after the PCC process. Fig.~\ref{fig_KEN} (a) shows the points of the maximum $|$PCC$|$ results to the guess keys during another $basemul$ call, where \( D = 15 \) and the number of candidates after PCC is equal to 4.  It can be seen that the correlations of four coefficients $564$, $1687$, $2010$, and $2264$ stand out from the crowd of guessed coefficients when setting the threshold to $0.930$ and their correlation values are very close. The phenomenon can lead to an inability to distinguish the correct one. This is another kind of false positive. If the attacker just picks one of the four candidates at random, the accuracy is only about $25$\%, which is still regarded as a failure. Therefore, we use the method mentioned in Section~\ref{AK} to compensate for such a case.

First of all, we compute the Kendall coefficient matrix of these four candidate coefficients. It should be noted that the time of this computation is negligible when compared to the complete PCC computing. Fig.~\ref{fig_KEN} (b) shows the values of the four maximum $|$Kendall’s tau$|$ when \( D = 15 \). It shows that Kendall’s tau expands the differences between the correct coefficient and false positives. As the number of power traces increases, the value of the correct coefficient tends to be more stable while the $|$Kendall’s tau$|$ of false positives decrease quickly. We can easily distinguish the correct key $564$ when \( D = 15 \) and output it as the final result. In other words, an attacker can take the correct coefficient $564$ for the $basemul$ call using at most 15 power traces. Similarly, we reduce the number of power traces to $11$ step by step and implement the PCC and Kendall’s tau computations. The corresponding results are shown in Fig.~\ref{fig_KEN} (c) and (d). It can be seen that the correct key $564$ can almost not be recognized. We can pick out the final correct candidate from those four candidates according to the results of the summations of their two kinds of corrections. Note that if we continue reducing the number of power traces, we may not be able to figure it out anymore. 

\subsubsection*{\textbf{Step2: Trial-and-Error Lattice Attack}}
We conduct the trial-and-error lattice attack for a 128-value group of $\mathbf{s}$ (hereinafter referred to as SK-128) using $16$ threads on a Xeon Gold 6342 @ $2.8$ GHz. The parameter of the $block$ size in the BKZ reduction is set to $50$ and the value of $max\_loops$ to $8$. We collect the CPA results of the first sixty $basemuls$ calls. Fig.~\ref{fig_reslut} shows the relationships between the success rate/time and the number of power traces represented in the blue curve and groups of histograms, respectively. The success rate is evaluated by repeatedly conducting the proposed attack many times and counting the success times. It can be seen that when the number of power traces is no less than $15$, all the experimental attacks are successful. Those cases with trace numbers smaller than $15$ can be compensated to some degree by running more trials of the lattice attack. Meanwhile, according to the decomposed time histograms in Fig.~\ref{fig_reslut}, the time of the proposed lattice attack heavily relies on the number of power traces, while the CPA is almost unchanged although the trace number is doubled. The main reason could be the different utilization ratios of the multi-threading server. Note that the CPA computations are independently executed with the different power traces while the trial computations of the lattice attack of SK-128 are executed in serial.

Table~\ref{attack time} presents the time of implementing the proposed attacks on Kyber at different security levels, using 15 power traces across twenty experiments. Kyber-512, -768, and -1024 have 4, 6, and 8 SK-128s, respectively, with each SK-128 being independent of the others. This allows us to recover the different SK-128s in parallel. The attack method we propose can recover the full secret key on different security levels of Kyber, with the success rate remaining consistent. The reason for the slight increase in time taken at the high security level is the high parallelism of Step1 and the duration of Step2 fluctuates around 5 minutes. It is worth noting that reducing the parallelism will increase the time required for Kyber-768 and Kyber-1024, but the success rate remains unaffected.

\begin{table}[!t]
\centering
\caption{Time Consumed to the Proposed Attacks on Kyber by Using 15 Power Traces in Twenty Experiments ($n=60$)}
\label{attack time}
\begin{tabular}{cccc@{\hspace{0.2mm}}c@{\hspace{0.2mm}}c@{\hspace{2mm}}c} 
\toprule
\multirow{2.5}{*}{\makecell{Security \\ Level}} & \multicolumn{1}{c}{Step1} & \multicolumn{1}{c}{Step2} & \multirow{2.5}{*}{\makecell{Total Time \\ ($min$)}} \\ 
\cmidrule(r){2-2} \cmidrule(lr){3-3}
          & Time / Thread   & Time / Thread &  &  \\ \midrule
SK-128  & 4.76 / $16\times1$ & 4.67 / $1\times1$ & 9.43  \\
Kyber-512 & 5.01 / $16 \times 4$ & 4.66 / $1\times4$ & 9.67  \\
Kyber-768 & 5.14 / $16 \times 6$ & 5.00 / $1\times6$ & 10.14  \\
Kyber-1024 & 5.30 / $16 \times 8$ & 5.33 / $1\times8$ & 10.63  \\ \bottomrule
\end{tabular}
\end{table}

\begin{table}[!t]
    \centering
    \caption{SCAs on Kyber to Recover SK-128}
    \label{tab:side-channel-attacks}
    \setlength{\tabcolsep}{0.5pt} 
    \begin{threeparttable} 
    \begin{tabular}{>{\centering\arraybackslash}p{0.8cm} > {\centering\arraybackslash}p{2.0cm} >{\centering\arraybackslash}p{1.5cm} >{\centering\arraybackslash}p{1.4cm} >{\centering\arraybackslash}p{1.7cm} >{\centering\arraybackslash}p{1.2cm}}
        \toprule
        Work & \makecell{Attack \\ Type} & \makecell{No Cip \\ Restriction} & \makecell{Attack \\ Target} & \makecell{Traces Num \\ for a Call} & \makecell{Num of \\ Calls}\\
        \midrule
        \cite{xu2021magnifying} & Pro SPA &  $\times$ & INTT, ME & 8$\sim$960 & 128\\
        \cite{mu2022voltage} & Pro TA & $\times$ & MR & 11 & 128\\
        \cite{sim2022chosen} & Non-Pro SPA & $\times$ & MR & 6$\sim$12 & 128\\
        \cite{yang2023chosen} & Non-Pro CPA & $\times$ & CM & 20$\sim$30 & 128\\
        \cite{kuo2023lattice} & Non-Pro CPA & $\checkmark$  & CM  & 800 & 50$\sim$87\\
        Ours & Non-Pro CPA & $\checkmark$ & CM & 11$\sim$15 & 60\\
        \bottomrule
    \end{tabular}
    \begin{tablenotes}    
    \footnotesize     
    \item Cip: Ciphertext, Num: Number, Pro: Profile, ME: Message Encoding, \\MR: Modular Reduction, CM: Coefficient Multiplication.   
    \end{tablenotes}           
    \end{threeparttable}
\end{table}

Table~\ref{tab:side-channel-attacks} shows the comparison results of SCAs on Kyber to recover SK-128. The attack types include the profiled SPA~\cite{xu2021magnifying}, profiled TA~\cite{mu2022voltage}, non-profiled SPA~\cite{sim2022chosen}, and non-profiled CPA~\cite{yang2023chosen,kuo2023lattice}. Our method belongs to the non-profiled CPA, which does not cost extra training step. Moreover, we only adopt random ciphertext, and thus we do not need extra effort like the oracle-based assistance to apply to the KEM scheme~\cite{ravi2024side}. The restriction ciphertext CPA requires oracle-based assistance to be applied to the KEM scheme. For the performance, as shown in the table, the total number of required power traces of ours is $[11,15]\times 60= [660,900]$, much smaller than the state-of-the-art work~\cite{yang2023chosen} which requires at least $20\times 128=2560$ power traces. If we slightly increase the traces number of the guess, the required time can be reduced to approximately $5$ minutes, much faster than the $20$ minutes recorded in~\cite{kuo2023lattice}.

\section{Conclusion}
\label{Sec:Con}
In this brief, we propose an efficient two-step attack for Kyber, combining an enhanced CPA and a trial-and-error lattice attack. 
Experimental results show that the proposed attack can accurately recover the full secret key of Kyber-512, -768, and -1024 in about $9\sim 10$ minutes with about 15 power traces. Moreover, the core idea of the proposed attack is a general methodology and can be easily extended to other lattice-based cryptography. It is worth noting that our proposed attack method can be applied to implementations on hardware like FPGAs by slightly modifying the leakage model. To mitigate these attacks, algorithm-level protections such as shuffling~\cite{xu2025hardware} and masking~\cite{bos2021masking} can be employed to protect the secret key, though these methods may involve trade-offs in efficiency and effectiveness.

\end{document}